\documentclass[aps,prl,twocolumn,groupedaddress]{revtex4-1}
\usepackage[normalem]{ulem}
\usepackage{graphicx}
\usepackage{epstopdf}
\usepackage{amssymb}
\usepackage{url}
\begin{document}

\title{Resolving Ultra-Fast Spin-Orbit Dynamics in Heavy Many-Electron Atoms}

\author{Jack Wragg}
\email[]{jack.wragg@qub.ac.uk}
\author{Daniel D. A. Clarke}
\author{Gregory S. J. Armstrong}
\author{Andrew C. Brown}
\author{Connor P. Ballance}
\author{Hugo W. van der Hart}
\affiliation{Centre for Theoretical Atomic Molecular and Optical Physics, School of Mathematics and Physics, Queen's University Belfast, Belfast, Northern Ireland, BT7 1NN}

\date{\today}

\begin{abstract}
We use R-Matrix with Time-dependence (RMT) theory, with spin-orbit effects included, to study krypton irradiated by two time-delayed XUV ultrashort pulses. The first pulse excites the atom to 4s$^{2}$4p$^{5}$5s. The second pulse then excites 4s4p$^{6}$5s autoionising levels, whose population can be observed through their subsequent decay. By varying the time delay between the two pulses, we are able to control the excitation pathway to the autoionising states. The use of cross-polarised light pulses allows us to isolate the two-photon pathway, with one photon taken from each pulse.
\end{abstract}


\maketitle

\section{}

Since the genesis of attosecond light pulses, nearly 20 years ago, great strides have been made in understanding how electronic motion can be observed and manipulated in a variety of atomic and molecular processes \cite{Krausz2009,Calegari2016}. Most of this understanding has been developed for systems involving light atoms, where the influence of relativistic interactions such as spin-orbit coupling is assumed to be sufficiently weak to be negligible over the timescales of interest. However, a number of recent advances have opened opportunities of harnessing spin-orbit effects to enable a deeper understanding of electron dynamics, including the observation of spin-orbit dependent effects in sub-cycle electron dynamics in ionisation from atoms by a near-infra-red pulse \cite{Sabbar2017}, effects in attosecond-timescale ionisation \cite{Wirth2011}, and time-delays in photoemission \cite{Jordan2017}.

This deeper understanding of spin-orbit interactions may also enhance our ability to manipulate the electron dynamics. For example, it has been shown that in ionisation from atoms, the spin polarisation of the ejected electrons can be controlled using circularly polarised light fields in conjunction with the spin-orbit interaction \cite{Hartung2016}. Moreover, many important biological processes such as photoreception and photosynthesis depend on population transfer between spin multiplicities mediated by the spin-orbit interaction \cite{Rettig1986}.

Despite these advances, the need for theoretical methods that are able to provide a complete description of spin-orbit interaction dynamics persists. Some steps have already been taken in this direction. For example, it has been shown that it is possible to account for spin-orbit effects using modified single-active-electron models \cite{Ivanov2014}, and analytic techniques such as the strong-field approximation \cite{Milovsevic2016}, PPT theory \cite{Barth2013} and the analytical R-matrix method \cite{Kaushal2018}. One of the most advanced methods is the time-dependent configuration-interaction singles approach \cite{Pabst2014}. However even this does not fully incorporate all characteristics of the spin-orbit interaction, as only its angular aspect is considered from first principles, with energy effects accounted for through empirical energy shifts \cite{Kaushal2018, Pabst2014}.

As a candidate for the full study of spin-orbit interaction, we propose the use of R-Matrix with Time-dependence (RMT) theory. Over the last decade, RMT has been established as one of the premier methods for the investigation of ultrafast dynamics in multi-electron systems. This theory builds upon the standard R-matrix theory for scattering, which was amended for the treatment of heavier atomic systems through the development of the Breit-Pauli R-matrix approach \cite{Berrington1995}. In this report, we demonstrate that by adopting the Breit-Pauli R-matrix approach within time-dependent R-matrix theory, we can extend RMT to describe ultrafast processes in heavy atoms which include relativistic effects.

\begin{figure}
 	\includegraphics[width=0.45\textwidth]{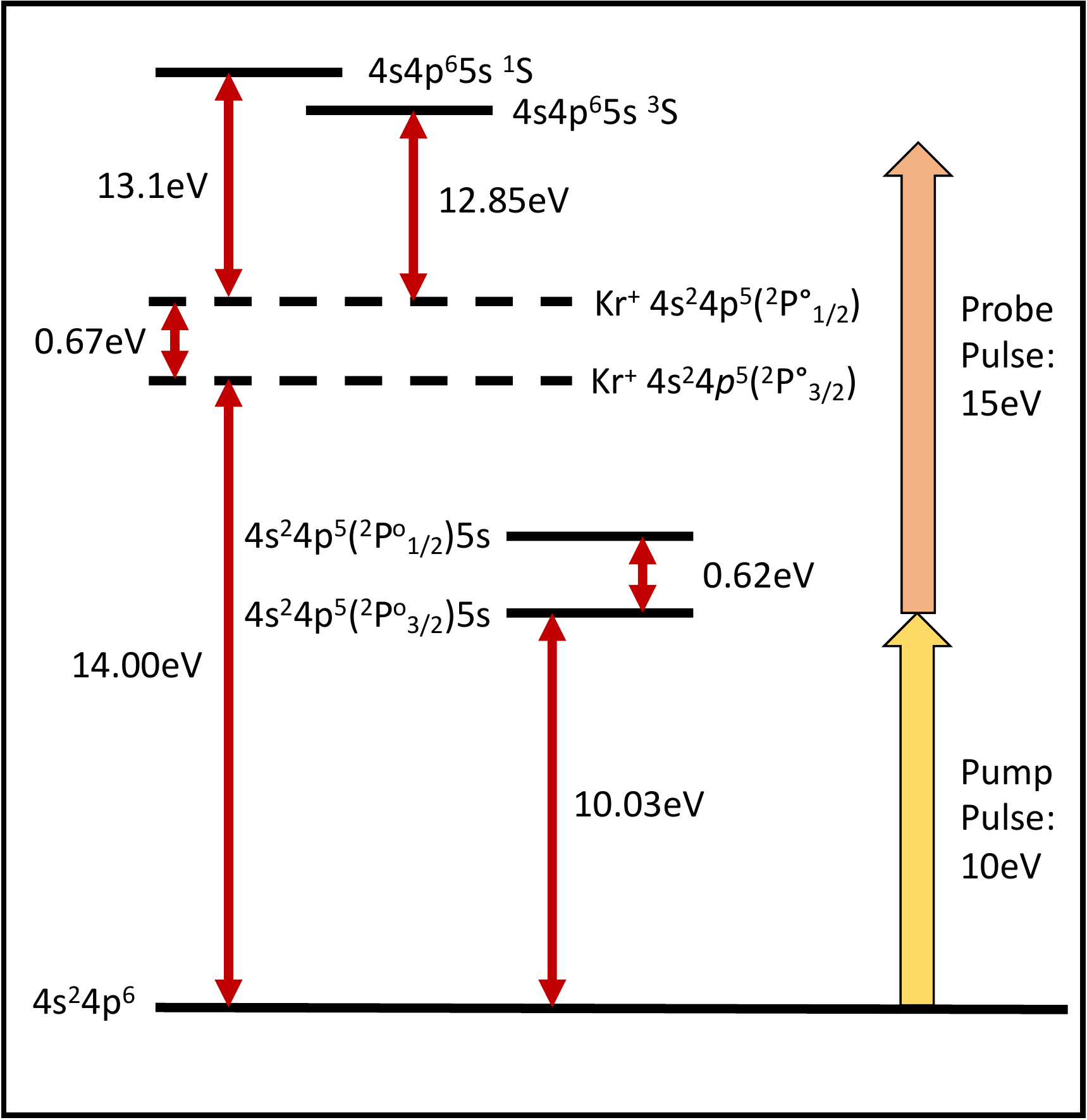}
 	\caption{\label{Levels} Energy levels of atomic krypton relevant to the autoionisation process shown in figure \ref{Spectra} (not to scale).}
 \end{figure}

We demonstrate the capability of this semi-relativistic RMT approach through application to an example experimentally-relevant problem: the combined effect of two time-delayed ultrafast laser pulses to induce two-photon excitation of an autoionising state in Kr (see Figure \ref{Levels}). Here, the first laser pulse (duration 6 cycles, photon energy 10 eV, and peak intensity $10^{13}$ W/cm$^{2}$) excites the Kr atom from the 4s$^{2}$4p$^{6}$ $^{1}$S ground state to 4s$^{2}$4p$^{5}$5s $^{1}$P$^{o}_{1}$. Then, spin-orbit coupling within the 4p$^{5}$ core will transfer population to the 4s$^{2}$4p$^{5}$5s $^{3}$P$^{o}_{1}$ state and back on a timescale given by the splitting between the levels within 4s$^{2}$4p$^{5}$5s. The second time-delayed laser pulse (duration 6 cycles, photon energy 15 eV, and peak intensity $10^{12}$ W/cm$^{2}$) then excites the Kr atom from 4s$^{2}$4p$^{5}$5s to 4s4p$^{6}$5s $^{1}$S$_{0}$ and $^{3}$S$_{1}$ states, which will subsequently autoionise. By varying the time delay between the pulses, we can control the excitation to these autoionising states. For both 6 cycle pulses, we employ a sin$^{2}$ profile, indicating 3 cycles of ramp on immediately followed by 3 cycles of ramp off. We use a cosine function for the carrier frequency of each pulse, such that unphysical effects resulting from an overall non-zero pulse displacement (for example as described in \cite{IgorIvanov2014}) are avoided.

In pump-probe schemes such as this, ionisation can typically occur through a vast number of pathways. Any number of photons can be absorbed from either pulse, and all paths interfere with one another. This interplay of interfering pathways can obscure the physics of direct interest, and has manifested itself in previous work (e.g. figure 6 in \cite{lysaght2009}, where a fast oscillatory change in ionisation is superimposed on the ionisation yield of interest). To bring the physics of interest to the fore, it is thus important to find schemes that isolate specific pathways. A benefit of the use of RMT is that it accounts for the combined effect of all pathways by design, so it is possible to investigate which factors enhance the signal from pathways of interest.  In this study, we examine the effect of varying the polarisation direction of the second pulse. We find this affords selectivity in this combined excitation-ionisation scheme, enabling the resolution of processes that might otherwise interfere and hide the signal from the physics of interest. Furthermore, we find that this use of cross polarised pulses allows the results to be interpreted in terms of the dynamics within the electron hole, driven by the spin-orbit interaction.

This work employs a similar scheme to that previously developed by W\"orner et al. \cite{Worner2011}, which described the effect of spin-orbit interaction induced dynamics associated with a superposition of P$_{1/2}$ and  P$_{3/2}$ states in noble gas ions. These dynamics can then be observed through measurement of the ionisation rate under a further ionisation step. Later works were able to observe these effects experimentally \cite{Fleischer2011,Fechner2014}. Here, our use of RMT theory allows the inclusion of a greater number of possible processes in the model, broadening the range of the physics we are able to study. Specifically, in the present work we modify the scheme to involve an autoionising state (previously studied experimentally , allowing the physics of interest to occur within the neutral krypton atom without requiring a further ionisation step. The autoionising state of interest (i.e. 4s4p$^{6}$5s) has been previously studied experimentally  from a time-independent perspective \cite{Baxter1982}.

We use the RMT method (for a full description see \cite{Moore2011} and \cite{Clarke2018}) to solve the time-dependent Schr\"odinger equation
\begin{equation}
	i\frac{\partial}{\partial t}\Psi(\mathbf{X},t)=\left[\hat{H}_{A}+\hat{H}_{SO}+\hat{H}_{D}(t)\right]\Psi(\mathbf{X},t),
\end{equation}
where $\mathbf{X}$ indicates spin and spatial dimensions for all electrons, $\hat{H}_{A}$ is the field-free atomic Hamiltonian, and $\hat{H}_{SO}$ is the Breit-Pauli spin-orbit interaction applied to all electrons. $\hat{H}_{D}(t)$ is the time-dependent dipole operator which describes the interaction of the laser with all electrons.
To implement $\hat{H}_{SO}$ we require semi-relativistic atomic data input, which we obtain from the fully ab initio, and widely used RMatrixI package \cite{Berrington1995,Ramsbottom2018,Smyth2018}. Hence all aspects of the spin-orbit interaction are taken into account, including the effect on both the angular and radial aspects of the wavefunction.

We build a model krypton atom, in which we include the 4p$^{-1}$ and 4s$^{-1}$  residual ion configurations. Bound orbitals are taken from pre-existing Hartree-Fock calculations \cite{Clementi1974}, and we represent the continuum orbitals with a basis of 50 B-splines of order 9, extending up to the inner region boundary at $20a_{0}$.  We obtain a Kr model with a ground state spin-orbit splitting in the residual ion of $0.668$ eV (experiment gives 0.666 eV \cite{Saloman2007}), and ionisation potential from the krypton atom ground state of $13.42$ eV (experiment gives 14.00 eV \cite{Saloman2007}). This model contains 29 electron emission channels when the spin-orbit interaction is omitted, and 66 channels when it is included. We retain all symmetries up to total orbital angular momentum $L=8$ when the spin-orbit interaction is omitted, and total angular momentum $J=8$ when included. For the cross-polarised pulses study, we include symmetries up to $M_{J}=3$, increasing the number of channels to 800. We propagate the wave function up to a final time of 47fs (the duration of each pulse is 2.44 fs), allowing plenty of time for ionised wavepackets to escape the atomic core. We use a time-step of 0.2 attoseconds, and describe the ejected electron up to a distance of 2168 $a_{0}$ from the core.

 \begin{figure}
 	\includegraphics[width=0.45\textwidth]{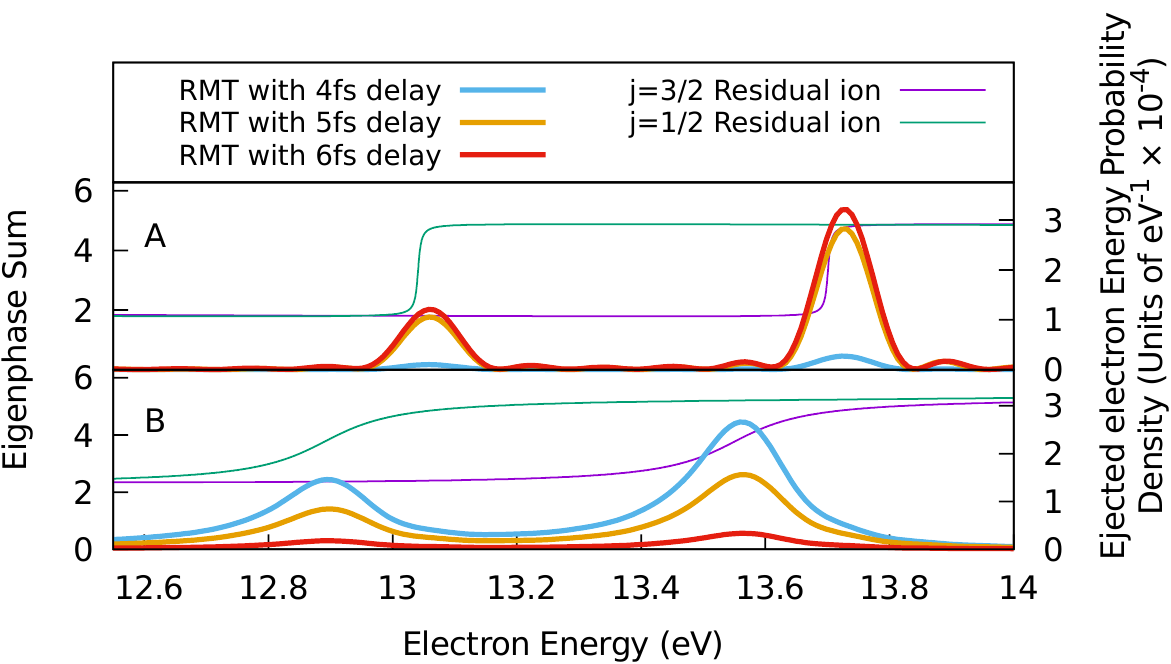}
 	\caption{\label{Spectra} Ejected electron spectra for delays of 4 fs, 5 fs and 6 fs between the two pulses. We show results for when the pulses are parallel polarised (frame A) and cross-polarised (frame B). We also show the corresponding eigenphase sum calculated using the same Kr model using the RMatrixI time-independent codes. In the upper frame we plot the eigenphase sum for the $J=0$ even symmetry, and in the lower frame we show that for $J=1$ even symmetry, to indicate the position of the resonances for parallel- and cross- polarised pulses respectively.}
 \end{figure}
 
 Figure \ref{Spectra} shows ejected electron spectra in the region corresponding to decay from the relevant autoionising state, for cases where the two pulses are parallel-polarised (frame A) and cross-polarised (frame B). We show spectra calculated using time-delays of 4, 5 and 6 fs between the two pulses. In all spectra we see two peaks, which can be explained by the spin-orbit splitting of the 4s$^2$4p$^5$ residual ion state: the 4s4p$^6$5s autoionising state decays either to 4s$^2$4p$^5$ $^2$P$^{o}_{1/2}$ or $^{2}$P$^{o}_{3/2}$. This is supported by the energy gap between the two peaks matching the splitting of these two states (0.66 eV).
 
For comparison, in Fig. \ref{Spectra}, we also include a time-independent Kr$^{+}$ + e$^{-}$ electron scattering calculation using the same krypton model. Here, we plot two versions of the eigenphase sum such that the peaks correspond to the decay signal from resonant states (i.e. we shift the spectra corresponding to decay to $j=\frac{1}{2}$  by 0.66 eV).  As parallel-polarised pulses will excite to $^{1}$S, and cross-polarised pulses will excite to $^{3}$S (as discussed below), we plot the eigenphase sum in the $J=0$ even symmetry to compare with the parallel-polarised spectra, and we plot the $J=1$ even symmetry to compare with data obtained from cross polarised pulses. We see that all peaks in the RMT spectra can be matched with the 4s4p$^{6}$5s autoionising resonances found in the time-independent calculation. The small difference in energy between the RMT and time-independent peaks (approximately $0.025$ eV) is ascribed to the potential energy experienced by the RMT wavepackets, as a residual field persists even at these large distances. 
  
From the spectra in figure \ref{Spectra}, we can obtain a measure of the population of the autoionised state by integrating over the peaks of interest. Inevitably this will introduce dependence on calculation length, as it is impractical to propagate to a time where the autoionising state can be considered to be fully decayed. While it might be possible to obtain quantitative results through fitting to an exponential decay, in this work we are most interested in the short-time dynamical effect of the time delay between the pulses rather than the absolute population autoionised. Thus we simply report the weight of the peak as measured at the end of the calculation.
 
 \begin{figure}
 	\includegraphics[width=0.45\textwidth]{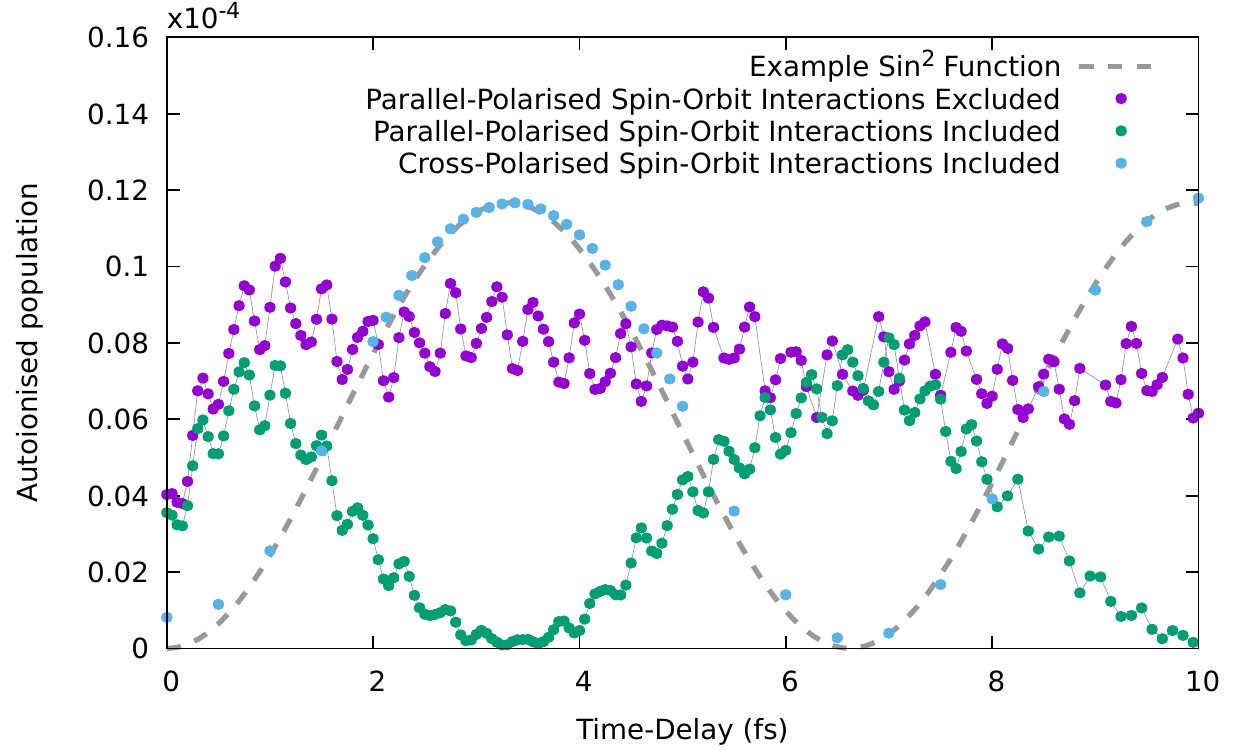}
 	\caption{\label{Results} Measure of ionisation resulting from the decay of the 4s4p$^{6}$5s state excited by a 6-cycle 10-eV pulse followed by a 6-cycle 15-eV pulse after a specified time-delay. The population is calculated at a time of 47fs after the start of the first pulse. We plot results for parallel-polarised cases where the spin-orbit interaction is omitted, and parallel-polarised and cross-polarised cases for when the spin-orbit interaction is included. We also plot a $\sin^{2}$ function (dashed grey line) of the frequency associated with the $0.615$ eV splitting, of amplitude scaled to fit the cross-polarised dataset.}
 \end{figure}

One of the most striking features of the results in Fig. \ref{Spectra} is the dependence of the weight of the peak on the time delay between the two pulses. We expand this result in Fig. \ref{Results} to plot the autoionised population (i.e. the area under the peaks) for a scan over a wider range of time delays. We first focus attention on the case where both pulses are polarised in the same direction. To show the effect of the spin-orbit interaction, we calculate datasets where the spin orbit interaction is included and omitted. On the longer timescale, the spin-orbit interaction introduces a strong periodic modulation on the degree of autoionisation from the atom. When the spin-orbit interaction is omitted, however, this periodicity disappears. We attribute this periodic behaviour to the splitting of the 4s$^{2}$4p$^{5}$5s state into the $^{2}$P$^{o}_{1/2}$ and $^{2}$P$^{o}_{3/2}$ core states. To support this explanation, we note that this longer oscillation has a period corresponding to approximately $0.62$eV (demonstrated by the corresponding grey dashed line), matching the splitting between the 4s$^{2}$4p$^{5}$($^{2}$P$^{o}_{1/2})$5s and  4s$^{2}$4p$^{5}$($^{2}$P$^{o}_{3/2})$5s levels with $J=1$ \cite{Saloman2007}.

  \begin{figure}
 	\includegraphics[width=0.45\textwidth]{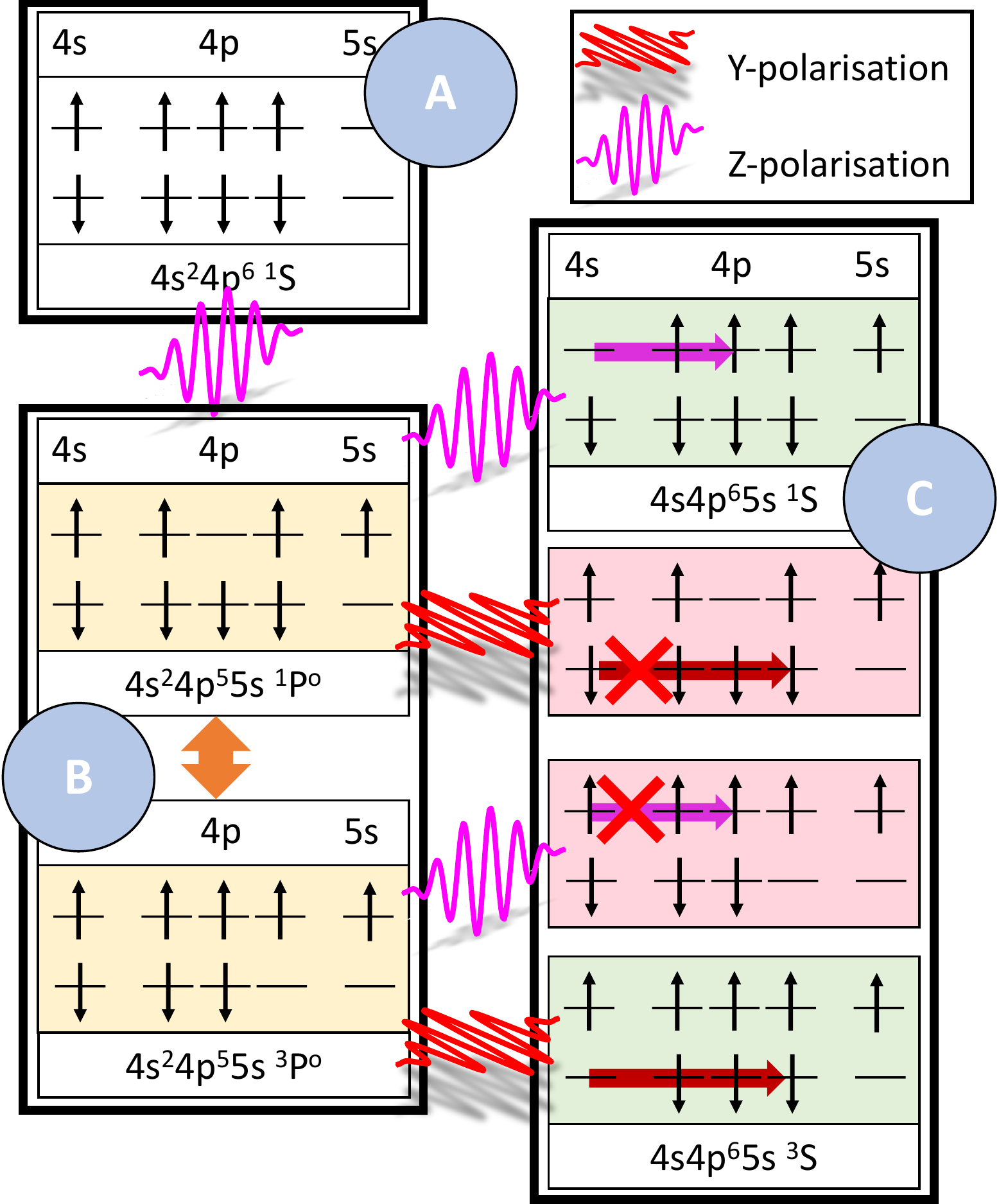}
 	\caption{\label{Explanation} A schematic to demonstrate the electron dynamics displayed in figure \ref{Results} in an uncoupled electron basis, shown in three stages (large blue circles). In stage A, the ground state krypton atom is excited by a 10 eV pulse. Stage B shows the effect of the spin-orbit interaction upon the 4p hole, and stage C shows the effect of the second 15 eV pulse for parallel-polarised and cross-polarised pulses. See text for more detail.}
 \end{figure}
 
To further explain this oscillation, we present similar results obtained using cross-polarised pulses in Fig. \ref{Results}. These results show an identical periodicity, but are out of phase with the results obtained using  parallel polarised pulses. To explain this phase difference, we consider these results in an uncoupled electron basis and the corresponding selection rules in Fig. \ref{Explanation}. Here the dynamics are considered in three stages. Stage A shows the ground state of the krypton atom, in which a 4p$_{0}$ electron is excited to 5s$_{0}$ by a pulse polarised in the $z$ direction (following the selection rule on the electron magnetic quantum number $\Delta m_{\ell}=0$). Stage B then shows the effect of the spin-orbit interaction, which moves the hole from 4p$_{0}$ to 4p$_{\pm1}$ and back again. Thus, by varying the time-delay we control the location of the hole state which is probed by the second pulse.
 
 Stage C shows the effect of the second pulse, where a 15 eV photon attempts to excite a 4s electron to 4p, and form the 4s4p$^{6}$5s autoionising state. The success of this process depends on whether the transition is allowed given the location of the hole and the selection rules of the second pulse. For example, a $z$-polarised pulse ($\Delta m_{\ell}=0$) will excite the 4s$_{0}$ electron to 4p$_{0}$ only when there exists a 4p$_{0}$ hole. If this is the case, then the autoionising state will form and its decay can be observed. If, however, there is no 4p$_{0}$ hole, the autoionising state cannot be reached.  Similarly, a $y$-polarised pulse ($\Delta m_{\ell}=\pm1$) can only excite the 4s$_{0}$ electron to a 4p$_{\pm1}$ hole. If this hole does not exist at the time of excitation, the autoionising state cannot be reached. 
 
With this perspective, the sinusoidal behaviour in figure \ref{Results} can be interpreted as representing the time-varying location of the 4p hole. When the parallel-polarised pulses show a maximum in the observed autoionisation decay, the hole can be considered to be in 4p$_{0}$. Correspondingly when the cross-polarised pulses show a maximum, the hole can be considered to be in 4p$_{\pm1}$.

In addition to the slow oscillation, both sets of parallel polarised data display a fast oscillation with time-delay which is absent in the cross-polarised data (we note that a numerical origin of this oscillation has been ruled out through careful checks). We attribute this oscillation to interference between three excitation pathways: The main pathway of interest which requires the absorption of a single photon from each pulse, and two other pathways, where two photons are absorbed from either of the pulses. 

We again consider the dipole selection rules, here in terms of $M_{J}$ (i.e. $\Delta M_{J}=0$ for z-polarised pulses, $\Delta M_{J}=\pm1$ for y-polarised pulses). When two photons are absorbed from the same pulse, insufficient time elapses between photon absorptions for the spin-orbit interaction to affect the spin of a core-electron. These excitation paths must thus result in excitation of the 4s4p$^{6}$5s $^{1}$S$_{0}$ state. In the parallel-polarised case therefore, the dipole selection rules ensure all pathways yield the same final state, and thus the pathways can interfere. However, in the cross-polarised case the pathway of interest (one photon delivered by each pulse) gives a  4s4p$^{6}$5s $^{3}$S$_{1}$  final state such that the pathway of interest no longer interferes with the others, and we do not observe the fast oscillation in the autoionisation yield. This explanation is supported by the frequency of the fast oscillation. While the oscillation does not seem to be purely sinusoidal, its dominant frequency is approximately $10$ eV, which corresponds to the energy gap between the first excited state and the ground state.

In the cross polarised case, we can thus isolate the excitation process of interest from the main interfering pathways. We note that previously published pump-probe results looking at spin-orbit effects seem to contain the same type of fast oscillations which could be attributed to similar interference effects (e.g. figure 1 in \cite{Fechner2014}). This shows how the wide range of processes included in the RMT approach enables a close representation of experiment, allowing these more subtle effects to be uncovered. Using the laser polarization to select specific pathways as we have done here could be realised as an effective method for resolving different processes in experiment. In this particular case, it also removes apparent ``noise'' which might otherwise obscure the signal of interest. We intend to further analyse this effect in a follow-up paper.

We fully expect that despite current technological limitations, this phenomenon should be observable in experiment. The autoionisation mechanism is mediated entirely by single photon processes, and thus limited laser intensity is not an issue. Furthermore, varying the relative pulse polarisation to isolate specific ionisation or excitation pathways does not face any technological barrier as far as we are aware. The most pressing limitation is the ability to create sufficiently short laser pulses in the XUV regime. An HHG source is the most promising candidate and it can be expected that the current drive towards XUV-pump/XUV-probe experiments will bring even the parameters discussed in this paper within reach \cite{Fushitani2013,Barillot2017}.

To summarise, we have developed an ab initio Breit-Pauli R-Matrix with Time-dependence approach and applied it to model ionisation from a krypton atom subject to two ultrafast laser pulses. We found that it is possible to control the $m_{\ell}$ of a core-hole at the moment of the second pulse by varying the time-delay between the pulses. The state of this core can be measured by exciting the atom to an autoionising state with the second pulse, and observing the resulting decay.

Through the demonstration of these effects, we have shown the importance of spin-orbit effects in ab initio calculations. While inclusion of the spin-orbit interaction can increase the computational resources required by the calculation, it is important that simulations of electron behaviour do not naively assume it is not required. Furthermore, we have shown how the Breit-Pauli approach can be combined with the RMT approach for arbitrary polarisation. Specifically, we demonstrate the potential for exploitation of spin-orbit effects in the observation and control of electron dynamics using parallel and cross-polarised excitation schemes.  These results give hope that further schemes to control electron dynamics at the level of the spin-orbit interaction may be possible. For instance, similar behaviour could be observed in more complex systems (e.g. molecules), where spin-orbit splitting occurs on similar energies, and hence dynamics occurs on similar timescales.

\begin{acknowledgments}
We acknowledge Jakub Benda and Zden\v ek Ma\v s\' in for their collaboration in developing and maintaining the RMT code. The data presented in this article may be accessed at Ref. \cite{DataDoi}. The RMT code is part of the UK-AMOR suite, and can be obtained for free at Ref. \cite{RMTurl}. This work benefited from computational support by CoSeC, the Computational Science Centre for Research Communities, through CCPQ. D.D.A.C. acknowledges financial support from the UK Engineering and Physical Sciences Research Council (EPSRC). A.C.B., H.W.v.d.H., G.S.J.A and J.W. acknowledge funding from the EPSRC under Grants No. EP/P022146/1, No. EP/P013953/1, and No. EP/R029342/1. This work relied on the ARCHER UK National Supercomputing Service \cite{Archerurl}, for which access was obtained via the UK-AMOR consortium funded by EPSRC.
\end{acknowledgments}

\bibliographystyle{apsrev4-1}
\bibliography{./Main}

\end{document}